\documentclass[eqsecnum,amsmath,preprintnumbers,superscriptaddress,nofootinbib,aps,preprint,12pt]{revtex4} 
\pdfoutput=1

\usepackage{amssymb,amsfonts,multirow}

\usepackage{epsfig}

\usepackage{placeins}
\usepackage{xspace}
\usepackage{cancel}

\usepackage{amssymb,url}
\usepackage{graphicx}
\usepackage{hyperref}

\usepackage{color}

\usepackage{array}
\usepackage{amsmath}
\usepackage{slashed}

\definecolor{rossoCP3}{cmyk}{0,.88,.77,.40}

\long\def\del #1 \enddel { }

\usepackage{graphicx}
\usepackage{amsmath}
\usepackage{amssymb}
\usepackage{subfigure}

\usepackage{epsfig}

\def\beq{\begin{equation}}
\def\eeq{\end{equation}}

\def\bea{\arraycolsep .1em \begin{eqnarray}}
\def\eea{\end{eqnarray}}
\def\Tr{{\rm Tr}}

\def\eps{\epsilon}

\def\al#1{\alpha_{#1}}

\def\s0#1#2{\mbox{\small{$ \frac{#1}{#2} $}}}
\def\0#1#2{\frac{#1}{#2}}

\def\grgl{\:\hbox to -0.2pt{\lower2.5pt\hbox{$\sim$}\hss}{\raise3pt\hbox{$>$}}\:}
\def\klgl{\:\hbox to -0.2pt{\lower2.5pt\hbox{$\sim$}\hss}{\raise3pt\hbox{$<$}}\:}

\newcommand{\wt}{\widetilde}

\def\lsim{\mathrel{\rlap{\lower4pt\hbox{\hskip1pt$\sim$}}
    \raise1pt\hbox{$<$}}}                
\def\gsim{\mathrel{\rlap{\lower4pt\hbox{\hskip1pt$\sim$}}
    \raise1pt\hbox{$>$}}}                

\newcommand{\drawsquare}[2]{\hbox{%
\rule{#2pt}{#1pt}\hskip-#2pt
\rule{#1pt}{#2pt}\hskip-#1pt
\rule[#1pt]{#1pt}{#2pt}}\rule[#1pt]{#2pt}{#2pt}\hskip-#2pt
\rule{#2pt}{#1pt}}

\newcommand{\Yfund}{\raisebox{-.5pt}{\drawsquare{6.5}{0.4}}}

\begin{document}

\title{ Supersymmetric asymptotic safety is not guaranteed }
\author{Kenneth Intriligator}
\email{keni@ucsd.edu}
\affiliation{\mbox{Department of Physics, University of California, San Diego, La Jolla, CA 92093 USA.}}
 \author{Francesco~Sannino}
\email{sannino@cp3-origins.net}
\affiliation{{\color{rossoCP3}CP${}^3$-Origins} \& the Danish Institute for Advanced Study, Danish IAS,
Univ.~of Southern Denmark, Campusvej 55, DK-5230 Odense}

\begin{abstract}
It was recently shown that certain perturbatively accessible, non-supersymmetric gauge-Yukawa theories have UV asymptotic safety, without asymptotic freedom: the UV theory is an interacting RG fixed point, and the IR theory is free.  We here investigate the possibility of asymptotic safety in supersymmetric theories, and use unitarity bounds, and the a-theorem, to rule it out in broad classes of theories. The arguments apply without assuming perturbation theory.  Therefore, the UV completion of a non-asymptotically free susy theory must have additional, non-obvious degrees of freedom, such as those of an asymptotically free (perhaps magnetic dual) extension.  
\\[1.5ex]
{Preprint: CP3-Origins-2015-036 DNRF90, DIAS-2015-036}

\end{abstract}

\maketitle

\newpage
\tableofcontents

\newpage
\section{Introduction}

The discovery of the Higgs crowns the Standard Model as one of the most successful theories of Nature.  Researchers have been desperately seeking hints of possible BSM
extensions of the Higgs sector.  Gauge-Yukawa theories are backbone of the Higgs sector, 
and it is therefore crucial to thoroughly investigate all possibilities for their dynamics.  

Recently, a surprise was found.  One can consider a gauge theory with too many matter fields to be asymptotically free in the UV, so it is instead infrared free.  Perturbation theory suggests that the theory is UV unsafe (the Landau pole), requiring a UV cutoff or completion.  A Yukawa interaction for the matter of this theory, on it's own, would also be IR-free and UV-unsafe.  Taken together, however, at least for a range of colors and flavors, the individually unsafe gauge and Yukawa interactions can cure each other and combine together to lead to a perturbatively accessible, fully interacting, non-supersymmetric, RG fixed point in the ultraviolet \cite{Litim:2014uca}.  See \cite{Pica:2010xq,Litim:2014uca} for the possibility that IR-free gauge and matter theory could have a UV fixed point even without the Yukawa interactions, albeit beyond perturbation theory. 

The phenomenon of UV asymptotic safety opens the door to new model building possibilities \cite{Sannino:2014lxa,Nielsen:2015una}, and provide a novel playground to explore, e.g.  the vacuum structure \cite{Litim:2015iea} and the thermodynamics of a new kind of matter \cite{Rischke:2015mea}. The theories are different  from  the time-honoured case of {\it complete asymptotic freedom} \cite{Gross:1973id,Politzer:1973fx,Gross:1973ju,Cheng:1973nv,Callaway:1988ya}, where the interactions instead shut off in the UV; see \cite{Holdom:2014hla,Giudice:2014tma} for recent studies. 

 Supersymmetry provides additional tools to explore the phases of gauge theories, beyond perturbation theory.  For example, in ${\cal N}=1$ duals, two UV asymptotically free theories can RG flow to the same IR SCFT, or an IR-free theory can be UV completed to an asymptotically free UV dual, with different gauge group and matter content \cite{Seiberg:1994pq} (see e.g. \cite{Intriligator:1995au} for a review).  In this latter case, the IR free theory avoids the Landau pole by completing to a different free theory in the UV; this is not the same as interacting asymptotic safety.

It is interesting to investigate if asymptotic safety can also occur in supersymmetric theories -- both for model building and for better understanding the phases of gauge theories. We will here rule out asymptotic safety for broad classes of supersymmetric theories, including ${\cal N} =1$ cousins  of the nonsupersymmetric asymptotically safe theories \cite{Litim:2014uca}.  Our methods and results do not rely on perturbation theory.  In some cases, a hypothetical asymptotically safe UV fixed point would violate unitarity bounds \cite{Mack:1975je}, and in other cases it would violate the 4d a-theorem \cite{Cardy:1988cwa,Osborn:1989td,Jack:1990eb,Cappelli:2000dv,Cappelli:2001pz,
Komargodski:2011vj,Komargodski:2011xv}, which can be explored in the susy context via the connection to 't Hooft anomalies for the superconformal R-symmetry \cite{Anselmi:1997am,Anselmi:1997ys,Intriligator:2003jj,Kutasov:2003iy,Kutasov:2003ux,Intriligator:2003mi,Barnes:2004jj,Kutasov:2004xu}.

The outline of this paper is as follows.  In Section \ref{SecPD}, we summarize the perturbatively accessible, non-supersymmetric asymptotically safe case \cite{Litim:2014uca}.   In Section \ref{Susynotes}, we 
summarize the main susy-based methods that we will use in the following sections.  In Section 
\ref{SQCDM}, we consider the direct ${\cal N} = 1$ cousin of the theory considered in \cite{Litim:2014uca}, based on ${\cal N}=1$ SQCD above the asymptotic freedom bound, $N_f>3N_c$.  We show that there cannot be a UV-interacting RG fixed point, neither with added gauge singlets Yukawa-coupled to the matter, nor for SQCD without the gauge singlets.  In Section \ref{Generalsusy} we apply susy-based methods to rule out asymptotically safe UV fixed points for theories with more general gauge group and matter content.  We offer our conclusions in Section \ref{Conclusions}.

\section{${\cal N}=0$ Asymptotic Safety, Brief Review of \cite{Litim:2014uca}}\label{SecPD}

Consider a massless theory with $SU(N_c)$ gauge group and $N_f$, $SU(N_c)$ fundamental, Dirac fermions $q_D$; we also write them as Weyl fermions $q$ and $\tilde q$.  The theory has a global $SU(N_f)\times SU(N_f)\times U(1)_B$ symmetry, and we include $N_f\times N_f$ complex scalar fields, which are gauge singlets.  The matter content is summarized in Table I.  The Lagrangian is  \bea
\label{F2}
L&=& 
- \s0{1}{4g^2} \Tr \,F^{\mu \nu} F_{\mu \nu}
+\Tr\left(
\overline{q}_D\,  i\slashed{D}\, q_D \right)
 +\Tr\,(\partial_\mu H ^\dagger\, \partial^\mu H) 
\nonumber 
\\
&&
+y \,\Tr\left(\wt{q} H q + {\rm h.c.} \right)
-h\,\Tr\,(H ^\dagger H )^2  
-v\,(\Tr\,H ^\dagger H )^2  \,,
\eea
and $\Tr$ indicates the appropriate trace over the suppressed color and flavor indices.
\begin{table}[b]
 \[ \begin{array}{|c|c|c c c |} \hline
{\rm Fields} &\left[ SU(N_c) \right] & SU_L(N_f) &SU_R(N_f) & U_V(1) \\ \hline 
\hline 
A_\mu & {\rm Adj} & 1 & 1 & 0  \\
 q &\Yfund &\overline{\Yfund }&1& 1   \\
\widetilde{q}& \overline{\Yfund}&1 &  {\Yfund}& -1    \\
 H  & 1 & \Yfund & \overline{\Yfund} & 0  \\
     \hline \end{array} 
\]
\caption{Field content of the  ${\cal N} =0$ theory and field. The $A_\mu$ are the gauge fields, $q$ and $\tilde q$ are Weyl spinors in the ($1/2,0$) Lorentz representation, and the $H$ are scalars. }
\label{QCD+Meson}
\end{table}
The classical theory is scale invariant with four marginal couplings: the gauge coupling $g$, the Yukawa coupling $y$, the quartic scalar couplings $h$ and the double-trace scalar coupling $v$.  In the quantum theory, it is convenient to introduce 
\beq\label{couplings}
\al g=\frac{g^2\,N_c}{(4\pi)^2}\,,\quad
\al y=\frac{y^{2}\,N_c}{(4\pi)^2}\,,\quad
\al h=\frac{{h}\,N_f}{(4\pi)^2}\,,\quad
\al v=\frac{{v}\,N^2_f}{(4\pi)^2}\,,
\eeq
with appropriate powers of $N_c$ and $N_f$ in the normalization to allow for the  Veneziano limit of large $N_c$ and $N_f$, holding fixed
\beq\label{xis}
x\equiv \frac{N_f}{N_c}\equiv \frac{11}{2}+\epsilon.
\eeq
The one-loop beta function is asymptotically free for $\epsilon <0$, and infrared free for $\epsilon >0$.  In this notation, the usual Banks-Zaks \cite{Banks:1981nn} limit is $\epsilon$ infinitesimally negative, whereas \cite{Litim:2014uca} instead considers $\epsilon$ infinitesimally positive.

The relevant beta functions $\beta _i (\alpha _g, \alpha _y, \alpha _h, \alpha _v)\equiv\partial_t\alpha_i$ for each coupling $i=(g,y,h,v)$ of the theory \eqref{F2}  have been obtained in \cite{Antipin:2013pya} in dimensional regularization using \cite{Machacek:1983tz,Machacek:1983fi,Machacek:1984zw,Jack:1984vj}. The point $\alpha _i=0$ is IR attractive, since none of the couplings are asymptotically free.  As shown in \cite{Litim:2014uca}, the beta functions vanish at non-zero couplings, compatible with  classical and quantum scalar potential stability \cite{Litim:2015iea}, given by 
 \beq\label{alphaNNLO}
\begin{array}{rcl}
\alpha_g^*&=&
\frac{26}{57}\,\eps+ \frac{23 (75245 - 13068 \sqrt{23})}{370386}\,\eps^2
+{\cal O}(\eps^3)
\\[1.ex]
\alpha_y^*&=&
\frac{4}{19}\,\eps+\left(\frac{43549}{20577} - \frac{2300 \sqrt{23}}{6859}\right)\,\eps^2
+{\cal O}(\eps^3)
\\[1ex]
\alpha_h^*&=&
\frac{\sqrt{23}-1}{19}\,\eps+{\cal O}(\eps^2)
\,,\\[1.ex]
\al {v}^*&=&
\frac{1}{19} (-2 \sqrt{23}+\sqrt{20 + 6 \sqrt{23}})\,\eps
+{\cal O}(\eps^2)\,.
\end{array}
\eeq
 The phase diagram of the theory was established in \cite{Litim:2014uca} at the next-to-leading order accuracy and extended  to the  next-to-next leading order  in \cite{Litim:2015iea}.  
 
 See Fig.~1 of  \cite{Litim:2015iea}  for the RG trajectories of the couplings $(\al g,\al y,\al h, \al v)$; the beta functions are all positive for $\alpha _{i=g,y,h,v}<\alpha _i^*$.  
 The interesting RG trajectory goes from the interacting fixed point at $\alpha _i=\alpha _i^*$ in the UV, and ends at the free theory, $\alpha _i=0$ in the IR. Along this one-dimensional line of physics, in the 4d $\alpha _i$ space, the Yukawa and scalar quartic couplings are all determined in terms of the running gauge coupling.  This dynamical relation among the couplings is dictated by the dimension of the critical surface.  In the UV, the gauge coupling approaches the interacting, asymptotically safe, UV fixed point by a power-law in the RG scale 
 \begin{equation}
\lim_{\mu/\mu_0 \to \infty} \alpha _g(\mu ) \to  \alpha _g^\ast + (\alpha _g(\mu_0) - \alpha _g^\ast) \left( \frac{\mu}{\mu_0}\right)^{-\tfrac{104}{171} \epsilon^2 + \mathcal{O}(\epsilon^3)},
\end{equation}
 (see \cite{Litim:2015iea} for the all-$\mu$ running, in terms of the Lambert function $W(\mu)$). 
 
It would interesting to extend the results beyond the perturbative regime via, for example,  first principle lattice simulations \cite{Hasenfratz:1991it,Gerhold:2007yb,Gerhold:2011mx,Chu:2015nha,Catterall:2013koa}. An alternative limit is QCD for fixed $N_c$ and large $N_f$ where, for the theory without scalars, at leading order $1/N_f$, an UV asymptotically safe fixed point seems also to appear \cite{Pica:2010xq,Litim:2014uca}.  Supersymmetry provides tools to explore beyond perturbation theory, as we will discuss in the following sections. 

\section{Supersymmetric RG fixed points and RG flows}\label{Susynotes}

We focus on theories in $d=4$ spacetime dimensions, with ${\cal N}=1$ supersymmetry.  A RG fixed point is an ${\cal N} =1$ superconformal field theory (SCFT), which necessarily has a conserved $U(1)_R$ global symmetry.  The $U(1)_R$ current is in the same supermultiplet \cite{Ferrara:1974pz} as the energy-momentum tensor and the supercharge currents; this leads to many useful exact relations.  For a unitary theory, the operators form unitary representations of the superconformal group, which implies that operator dimensions have various lower bounds.  For example, regardless of supersymmetry, all gauge invariant spin $j=\bar j=0$ operators have the lower bound (generators act with implicit commutators) \cite{Mack:1975je} (see also e.g. \cite{Grinstein:2008qk})
\beq\label{dimone}
D({\cal O})\geq 1, \qquad D({\cal O})=1 \leftrightarrow P_\mu P^\mu ({\cal O})=0,
\eeq
so the bound is saturated if and only if the operator ${\cal O}$ is a decoupled, free field.  Chiral primary operators have dimension, $D$, and superconformal $U(1)_R$ charge, $R$, related by 
\beq\label{chiralprimary}
D({\cal O})=\frac{3}{2} R({\cal O}).
\eeq
In particular, using \eqref{chiralprimary} for the matter chiral superfields $Q_i$ of a supersymmetric gauge theory relates the matter anomalous dimensions $\gamma _i$ to their superconformal $U(1)_R$ charge.  
\begin{equation}
D(Q_i)\equiv 1+\frac{1}{2}\gamma _i(g)=\frac{3}{2}R(Q_i) \equiv \frac{3}{2} R_i .\label{aba:eq5}
\end{equation}

The conformal anomaly $a$ of the SCFT is exactly given by the superconformal $U(1)_R$ 't Hooft anomalies  \cite{Anselmi:1997am,Anselmi:1997ys} (we rescale the overall normalization factor for convenience)
\beq\label{aanselmi}
a(R)=3{\rm Tr}U(1)_R^3-{\rm Tr}U(1)_R.
\eeq
For a gauge theory with gauge group, $G$, and matter fields $Q_i$, in representations $r_i$ of $G$, the 't Hooft anomalies evaluate to 
\beq\label{agaugetheory}
a(R_i)=2|G|+\sum _i |r_i| a_1(R_i),
\eeq
where $|G|=|r_{\rm Adjoint}|$ is the number of generators of the gauge group, $|r_i|$ is the dimension of the $r_i$ representation, $R_i\equiv R(Q_i)$ is the $U(1)_R$ charge of $Q_i$, and we define the function 
\begin{equation}
a_1(R)\equiv 3(R-1)^3-(R-1)~.\label{aRex}
\end{equation}

Among all possible, conserved $U(1)_R$ symmetries, the superconformal $U(1)_R$ is that which maximizes  $a(R)$ \cite{Intriligator:2003jj}.  For example, for a chiral superfield $X$ of charge $R(X)=R$ (so $R(\psi _X)=R-1$), the function is $a(R)=a_1(R)$ in \eqref{aRex}.  The function $a_1(R)$ has a local maximum at the free-field value, $R={2\over 3}$, and a local minimum at $R={4\over 3}$.  Indeed $a_1(R)=-a_1(2-R)$, so $a_1(R=1)=0$, fitting with massive operators contributing $a=0$.     Note also that $a_1(R)$ is below the local maximum, $a_1(R)<a_1(R=2/3)$, for all $R$ in the range $R<5/3$ (see \cite{Intriligator:2005if} for a related conjectured phase diagnostic).
 For unconstrained, i.e. free chiral superfields, we maximize the function \eqref{aRex} to get  $R_*=2/3$,  the free-field value of the R-charge, corresponding to $D (X)=1$.  With interactions, we maximize $a(R)$ subject to the constraint that the interactions do not violate the R-symmetry. Accidental symmetries affect a-maximization \cite{Kutasov:2003iy}(see also e.g. \cite{Intriligator:2003mi}), which if present leads to a larger value of $a$. 

Away from a RG fixed point, the beta functions are proportional to how the couplings break the superconformal $U(1)_R$.  The gauge coupling beta function is proportional to the ABJ triangle anomaly of the $U(1)_R$ current with two $G$ gauge fields, i.e. ${\rm Tr}~G^2U(1)_R$:
\beq\label{gaugebeta}
\beta (g)=-\frac{3g^3}{16\pi ^2}f(g^2){\rm Tr}~ G^2U(1)_R, \qquad 
{\rm Tr}~G^2U(1)_R=T(G)+\sum _i T(r_i)(R_i-1).
\eeq
Our normalization for the quadratic Casimir of the adjoint $T(G)$ is  $T(SU(N_c))=N_c$, so the fundamental representation of $SU(N)$ has $T(r_{\rm fund})=\frac{1}{2}$. The function $f(g^2)=1+{\cal O}(g^2)$ is scheme dependent (and presumed positive).  Using \eqref{aba:eq5} gives the statement of the NSVZ exact beta function (NSVZ also have a favored, specific scheme choice for $f(g^2)$) \cite{Novikov:1983uc}:
\beq\label{gaugebetaNSVZ}
\beta (8\pi ^2g^{-2})=f(g^2)(3T(G)-\sum _i T(r_i)(1-\gamma _i(g))).
\eeq
For superpotential terms $W_y$, the holomorphic coupling $y$ prefactor has beta function
\beq\label{betah}
\beta (y)=\frac{3}{2} y (R(W_y)-2).
\eeq

\section{Supersymmetric QCD with $N_f > 3 N_c$ is unsafe} \label{SQCDM}
As a first class of examples, we consider the ${\cal N} =1$ cousin of the theory \eqref{F2}, with superfield content and quantum symmetries summarized in Table~\ref{SQCD+Meson}. 
 \begin{table}[t]
\[ \begin{array}{|c|c|c c c  c|} \hline
{\rm Fields} &\left[ SU(N_c) \right] & SU_L(N_f) &SU_R(N_f) & U_V(1) & U(1)_R\\ \hline 
\hline 
W_\alpha & {\rm Adj} & 1 & 1 & 0 & 1  \\
 Q &\Yfund &\overline{\Yfund }&1& 1 & 1 - \frac{N_c}{N_f}  \\
\widetilde{Q}& \overline{\Yfund}&1 &  {\Yfund}& -1 & 1 - \frac{N_c}{N_f}  \\
 H  & 1 & \Yfund & \overline{\Yfund} & 0  &  2\frac{N_c}{N_f}\\
     \hline \end{array} 
\]
\caption{The ${\cal N}=1$ superfield content, cousins of the theory \eqref{QCD+Meson}. $W_{\alpha}$ is the gauge vector superfield, $Q$ and $\wt{Q}$ are the matter chiral superfields, and $H$ are gauge singlet chiral superfields.  }
\label{SQCD+Meson}
\end{table}
The superpotential is: 
\begin{equation}\label{WH}
W = y {\rm Tr} \,Q H \wt{Q} \ ,
\end{equation}
with $y$ the Yukawa coupling and ${\rm Tr}$ contracts the implicit gauge and flavor indices.  The quartic in $H$ interactions in \eqref{F2} are incompatible with holomorphy of the superpotential (and the $SU(N_f)\times SU(N_f)$ global symmetry does not allow for a holomorphic variant). 
%
%

Taking $N_c$ and $N_f$ large, and properly rescaling the couplings as 
\beq\label{couplings}
\al g\equiv \frac{g^2\,N_c}{(4\pi)^2}\,,\qquad
\al y\equiv \frac{y^{2}\,N_c}{(4\pi)^2}\,,
\eeq
the two loops beta functions are, dropping terms subleading in $1/N_c$, 
\begin{eqnarray}\label{SQCDbetas}
\beta(\alpha_g) & \approx  & - 2 \alpha_g^2 \left[3 - \frac{N_f}{N_c} + \left(6 - 4  \frac{N_f}{N_c} \right)  \alpha_g + 2 \frac{N_f^2}{N_c^2} \alpha_y  +{\cal O}(\alpha ^2)\right] \ , \nonumber \\ && \nonumber\\
\beta(\alpha_y) & \approx  &  2 \alpha_y\left[ \left(2\frac{N_f}{N_c}+ 1\right) \alpha_y - 2\alpha_g +{\cal O}(\alpha ^2)\right] \ .
\end{eqnarray}
To connect with \eqref{gaugebetaNSVZ} and \eqref{betah} note that they give (using \eqref{aba:eq5}, and $R(W)=R(H)+2R(Q)$)
\beq\label{sqcdbetas}
\beta (\alpha _g)=-2\alpha _g^2 f(\alpha _g) \left(3-\frac{N_f}{N_c}(1-\gamma _Q)\right), \qquad \beta (\alpha _y)=\alpha _y (\gamma _H+2\gamma _Q),
\eeq
and to the relevant order (dropping ${\cal O}(\alpha _{g, y}^2)$ and ${\cal O}(1/N_c)$) we have
\beq
f(\alpha _g)\approx 1+2\alpha _g, \qquad \gamma _Q\approx -2\alpha _g+2\frac{N_f}{N_c} \alpha _y, \quad \gamma _H\approx 2 \alpha _y
\eeq

For $N_f>3N_c$, the theory is not asymptotically free, and $\alpha _g=\alpha _y=0$ is IR-attractive.   We first note that there cannot be a perturbative, interacting, UV-safe fixed point. Define $\epsilon \equiv   N_f/N_c  - 3 $, with $N_f$ and $N_c$ such that $0 < \epsilon \ll 1$.   The condition $\beta (\alpha _y)=0$ then gives $\alpha_y \approx \frac{2}{7} \alpha_g $, and then  $\beta (\alpha _g)$, to leading order in small $\epsilon$, is 
\begin{equation}
\beta(\alpha_g) \approx  2 \alpha_g^2 \left[  \epsilon + \frac{6}{7} \alpha_g \right] \ ,
\end{equation}
so the relative sign is such that, for positive $\epsilon$, $\beta (\alpha _g)\neq 0$ unless $\alpha _g=0$. 

We now argue, including at the non-perturbative level, that there cannot be a UV-safe interacting SCFT.  We assume that the superconformal $U(1)_R$ is not emergent or accidental, in which case it must be the anomaly free R-symmetry that is preserved by the superpotential, i.e. the $U(1)_R$ given in Table II; this ensures that the beta functions \eqref{sqcdbetas} vanish. The dimension of the operators $H$ are then determined to be
\begin{equation}\label{DH}
D(H) = \frac{3}{2} R(H)=3 \frac{N_c}{N_f} \ . 
\end{equation}
These statements are all correct for the {\it IR} SCFT fixed point when $N_f<3N_c$.  But,  for $N_f>3N_c$, \eqref{DH} would violate the unitarity bound \eqref{dimone}.  This is impossible, since the original theory is unitary. There thus cannot be an interacting UV SCFT for $N_f>3N_c$.

It follows from Seiberg duality \cite{Seiberg:1994pq} that the $N_f>3N_c$ theory can actually be UV-completed to an asymptotically {\it free} dual, rather than an interacting, UV-safe, SCFT. 

A potential loophole in the above argument is that apparent unitarity bound violations mean that the corresponding field -- in this case $H$ -- is instead a free, decoupled field.  So we set $\alpha _y=0$, which is equivalent to considering SQCD without the $H$ singlets.   We now argue that SQCD without the $H$ fields also cannot have a UV safe RG fixed point for $N_f>3N_c$.  Consider first the perturbative regime, 
$\epsilon \equiv   N_f/N_c  - 3 $, with $0 < \epsilon \ll 1$.  Taking  $\alpha _y=0$ in \eqref{SQCDbetas} gives 
\begin{equation}
\beta(\alpha_g) \approx  2 \alpha_g^2 \left[  \epsilon + {6} \alpha_g \right] \ ;
\end{equation}
so $\beta (\alpha _g)\neq 0$ for  $\epsilon>0$ and $\alpha _g\neq 0$, i.e. there is no perturbative UV-safe fixed point.  

We now rule out asymptotic safety beyond perturbation theory.  First note that, for $\alpha _g\neq 0$, having $\beta (\alpha _g)=0$ requires that the superconformal $U(1)_R$ be that in Table II, so
\begin{equation}
D_{SCFT}(Q)  = D_{SCFT}(\wt{Q}) \equiv  1 + \frac{1}{2} \gamma _Q= \frac{3}{2} R_{SCFT}(Q)=\frac{3}{2} \left( 1 - \frac{N_c}{N_f}\right) .\ 
\end{equation}
The gauge invariant chiral operators, the mesons ${\cal M}=Q\tilde Q$ and baryons ${\cal B}=Q^{N_c}$ have
\begin{eqnarray}
D_{SCFT}({\cal M}) & = & \frac{3}{2}R_{SCFT}({\cal M}) = 3 \frac{N_f - N_c}{N_f} \ , \nonumber \\
D_{SCFT}({\cal B}) & = & D_{SCFT}(\wt{\cal B}) =  \frac{3}{2}R_{SCFT}({\cal B}) = \frac{3}{2} N_c \frac{N_f - N_c}{N_f} \ . 
\end{eqnarray}
These expressions are indeed correct for the IR fixed point of SQCD  in the conformal window \cite{Seiberg:1994pq} $\frac{3}{2}N_c< N_f<3N_c$.    Extrapolating to the 
hypothetical, UV fixed point for $N_f>3N_c$, these expressions would apply, and the operators would satisfy the unitarity bound \eqref{dimone}.  

We therefore rule out a hypothetical, interacting UV fixed point for $N_f>3N_c$ SQCD on different grounds, 
by noting that it would violate the 4d  $a$-theorem \cite{Cardy:1988cwa,Komargodski:2011vj,Komargodski:2011xv}, 
\beq\label{athm}
\Delta a \equiv a_{UV}-a_{IR}>0.
\eeq
If the hypothetical UV-safe fixed point exists, its $a$ would be given by 
 \eqref{agaugetheory} with the interacting, anomaly free $U(1)_R$, i.e. $R_{SCFT}(Q)=(N_f-N_c)/N_f$,
 \begin{equation}\label{auvwrong}
a_{\hbox{hypothetical UV}}= a_{\rm SCFT}=2 (N_c^2-1) + 2 N_f N_c a_1(R=1-\frac{N_c}{N_f}) \ ,
\end{equation}
with $a_1(R)$ the function \eqref{aRex}. 
There would be a RG flow from the hypothetical UV SCFT to the IR-free theory with $\alpha _g=0$ and thus $R(Q)=2/3$:
 \begin{equation}\label{afree}
a_{\rm IR}=a_{\rm free}= 2(N_c^2-1) + 2 N_f N_c a_1(R=2/3)= 2 (N_c^2 -1)+ \frac{4}{9} N_f N_c \ .
\end{equation} 
So this RG flow would violate the a-theorem  \eqref{athm}:
\begin{equation}\label{wrongdeltaa}
a_{\rm UV-safe}-a_{\rm IR-free}=2N_cN_f\left( a_1(R=1-\frac{N_c}{N_f})-a_1 (R=2/3)\right)<0,
\end{equation}
where the inequality is evident from graphing $a_1(R)$ in \eqref{aRex} since, for $3N_c<N_f<\infty$, the R-charge $R(Q)=1-\frac{N_c}{N_f}$ is in the range $\frac{2}{3}<R(Q)<1$, and $a_1$ in that range is below its local maximum $a_1(R=2/3)$.  Given the wrong sign \eqref{wrongdeltaa}, we conclude that there cannot be an interacting, UV safe fixed point.  Instead, 
IR-free electric SQCD theory can be UV-completed to an asymptotically free magnetic dual.  For SQCD in the conformal window, the identification of the endpoints is opposite from that in \eqref{wrongdeltaa} (free in the UV and interacting in the IR). Both those RG flows of course do satisfy the a-theorem \cite{Anselmi:1997am, Anselmi:1997ys}.

\section{Susy theories with general gauge group and matter}\label{Generalsusy}

We first consider susy gauge theories without superpotential terms, $W_{\rm tree}=0$.  The exact beta function for the gauge coupling is as in \eqref{gaugebeta}, and the condition $\beta (g)=0$ is equivalent to the condition that the superconformal $R_{\rm SCFT}(Q_i)\equiv R_{\rm{SCFT},i}$ is anomaly free:
\beq\label{gbetaagain}
\beta (\alpha _g)=0 \quad \leftrightarrow \quad T(G)+\sum _i T(r_i)(R_i-1)=0\quad \leftrightarrow\quad 3T(G)-\sum _i T(r_i)(1-\gamma _{Q_i})=0.
\eeq
\beq\label{notAF}
\hbox{So not asymptotically free has:}\qquad 3T(G)-\sum _i T(r_i)<0.
\eeq
It is then impossible to satisfy \eqref{gbetaagain} perturbatively, since all perturbative $\gamma _{Q_i}$ are negative. 

Generalizing the argument of the previous section, we can moreover rule out UV-safe SCFTs in the range \eqref{notAF}, without relying on perturbation theory.  The RG flow from the hypothetical, asymptotically safe SCFT in the UV, to the free $\alpha _g=0$ theory in the IR, would violate the a-theorem.  Using \eqref{agaugetheory}, the hypothetical flow has 
\beq\label{gaugegenda}
\Delta a = a_{\rm{UV-safe}}-a_{\rm{IR-free}}=\sum _{\hbox{all matter} \ Q_i} |r_i| \left(a_1(R_{{\rm SCFT}, i})-a_1(R_i=2/3)\right).
\eeq
The general expression for $R_{{\rm SCFT}, i}$ follows from a-maximization \cite{Intriligator:2003jj}, i.e. maximizing $a_1(R_{{\rm SCFT}, i})$ over the $R_i$, subject to the 
constraint in \eqref{gbetaagain}.  This gives \cite{Kutasov:2003ux},
\beq
R_{{\rm SCFT}, i}=1-\frac{1}{3}\left(1+\frac{\lambda T(r_i)}{|r_i|}\right)^{1/2},
\eeq
where the Lagrange multiplier is determined via \eqref{gbetaagain}.  In the asymptotically free case, this yields $\lambda = (g_*^2/2\pi ^2)+{\cal O}(g_*^4)$ (the higher order terms are scheme dependent) \cite{Kutasov:2003ux,Barnes:2004jj,Kutasov:2004xu}.  In the non-asymptotically free case \eqref{notAF}, on the other hand, the constraint leads to $\lambda<0$, and all $R_{SCFT, i}$ would be in the range $2/3<R_i<1$.  It is then clear from the graph of $a_1(R)$ that, since $R_i<5/3$, every term in  \eqref{gaugegenda} is negative, $a_1(R_{{\rm SCFT}, i})-a_1(R_i=2/3)<0$, 
the flow from the hypothetical, UV-safe fixed point would violate the a-theorem.   So susy gauge theories with $W=0$ cannot be interacting UV-safe, and IR free, without some new element (for example, some accidental symmetry in the interacting UV theory). 

We now consider adding superpotential terms.  The upshot of a-maximization is that the microscopic fields have superconformal $R$-charges given by \cite{Barnes:2004jj}
\beq\label{amaxRgen}
R(Q_i)=1-\frac{\epsilon _i}{3}\sqrt{1-2\gamma _i^{(1)}(\lambda)},
\eeq
where $\epsilon _i=\pm 1$ (see \cite{Barnes:2004jj,Amariti:2012wc} for curiosities related to $\epsilon _i$ sign changes in RG flows) and $\gamma _i^{(1)}$ are linear in the $\lambda$ and related to the one-loop anomalous dimensions.  The values of $R_{SCFT}(Q_i)$ are obtained from \eqref{amaxRgen} by solving for the $\lambda _{SCFT}$ such that all $R$-charge conservation constraints are satisfied, i.e. all gauge groups $G$ have ${\rm Tr}G^2U(1)_R=0$ and all $W_y$ have $R(W_y)=2$, i.e. all beta functions \eqref{gaugebeta} and \eqref{betah} vanish.   Note that 
all \eqref{amaxRgen} yield $R_{\rm{SCFT}, i}<4/3$, and thus all $a_1(R=R_{\rm{SCFT},i})<a_1(R=2/3)$.  The RG flow from a hypothetical UV-safe SCFT, to the IR-free theory, would have $\Delta a$ given by \eqref{gaugegenda}, which again would violate the a-theorem because every term in the sum has the wrong sign. 

Also, as in the examples \eqref{WH}, gauge singlets $H$ coupled to composite operators ${\cal O}$, of  classical dimension $D_{cl}({\cal O})=2$, leads to a unitarity bound problem for hypothetical UV-safe SCFTs.  Because the theories are not asymptotically free, the condition \eqref{gbetaagain} leads to $R_{\rm{SCFT}}({\cal O})>R_{\rm{free}}({\cal O})$, and then $R(W)=2$ requires $R(H)<2/3$; then $D_{\rm{UV-safe}}=\frac{3}{2}R(H)<1$, violating the unitarity bound.  The $H$ fields must instead remain free; this fits with the fact that there is no interacting, UV-safe SCFT after all. 

Note that ${\cal N}=4$ supersymmetric theories, deformed to ${\cal N}=1$ by making the cubic superpotential coupling, $y$, differ from the gauge coupling, $g$, have RG flows somewhat similar to, but crucially different from, asymptotic safety.  Setting either $y=0$ or $g=0$, the other coupling is {\it marginally} irrelevant.  With both $y\neq 0$ and $g\neq 0$, there are RG flows with an IR-attractive line of interacting SCFTs for all $g=y$; the UV limit of the flows for $g\neq y$ has Landau poles.  The adjoint chiral superfields $\Phi _{i=1,2,3}$ have $R(\Phi _i)=2/3$ for all $g=y$ SCFTs (the $\Phi_i$ are not free, because they are not gauge invariant).  These examples do not contradict our general arguments.  Their one-loop beta function in \eqref{notAF} are zero, and that there is a line of fixed points, rather than a RG flow, between the interacting $(g=y\neq 0$) and free ($g=y=0$) theories, all with equal value of $a$.  And there is no UV SCFT, only IR SCFTs.  There are many similar ${\cal N}=1$ SCFT examples, along the lines of \cite{Leigh:1995ep}.  

Another example of a non-asymptotically free theory is ${\cal N}=1$ gauge theory with the three adjoints of ${\cal N}=4$, plus $N_f$ flavors of matter in the fundamental  (as is sometimes considered in the context of AdS/CFT, via added D7 branes). By our general argument, such theories cannot\footnote{This fits with the dual gravity analysis in \cite{D'Hoker:2010mk}. KI thanks David Mateos for pointing out this reference.} have a UV-safe SCFT without violating the a-theorem.

\section{Conclusions}
\label{Conclusions}
We investigated the nonperturbative gauge dynamics of ${\cal N}=1$ super QCD both with, and without, gauge singlet (dual meson) fields $H$, in the $N_f$, $N_c$ regime where asymptotic freedom is lost.  Unlike the non-supersymmetric case \cite{Litim:2014uca}, the $H$ fields do not help to achieve asymptotic safety.  Instead, the $H$ fields would necessarily violate a unitarity bound in a hypothetical, asymptotically safe SCFT unless their Yukawa coupling is zero and they are decoupled.  We showed that the theory without the $H$ fields also cannot have a UV-safe SCFT, because it would violate the a-theorem.  We used a-maximization to show that the same issues arise for general gauge groups and matter content.  

We arrive at the interesting conclusion that there is a fundamental obstacle to supersymmetric asymptotic safety.  Any sensible, UV completion of unsafe IR theories must include many additional degrees of freedom, e.g. those of an unknown, asymptotically free dual.

\vskip 1cm
{\bf Note added in revised version:}  We would like to thank Steven Martin and James Wells for bringing their relevant paper \cite{Martin:2000cr} to our attention.   Although there is considerable overlap, our paper contains additional methods -- for example, using a-maximization, which had not yet been developed at the time  \cite{Martin:2000cr} was written.  Intriguingly, \cite{Martin:2000cr} suggested a possible way to construct superconformal UV fixed point theories via superpotential terms.  To quote an example from \cite{Martin:2000cr}, consider $SU(N_c)$ gauge theory with $N_f$ fundamental flavor chiral superfields $Q$, $\widetilde Q$, two adjoints chiral superfields $A_1$ and $A_2$, and a superpotential $W=A_1Q\widetilde Q+A_1^3$.  The theory is IR free for $N_f>N_c$.  A hypothetical UV interacting fixed point would have $R(A_1)=R(Q)=R(\widetilde Q)=2/3$, and $R(A_2)=(N_f+N_c)/3N_c$, which would satisfy $a_{UV}>a_{IR}$ if $R(A_2)>5/3$, i.e. if $N_f>4N_c$.  We subjected this example to a few additional consistency conditions, including those in this paper and also e.g. verifying that $a/c$ satisfies the inequalities of \cite{Hofman:2008ar}.  We have not yet found any inconsistency to definitively rule out the hypothetical UV-safe SCFT for $N_f>4N_c$.  We note however that replacing the adjoint $A_1$ with a gauge singlet $S_1$ would {\it not} give an interacting SCFT: in that case, since $S_1$ is a gauge invariant chiral operator, $R(S_1)=2/3$ would imply $\Delta (S_1)=1$ and then $S_1$ would necessarily be a free-field, incompatible with the superpotential coupling having an interacting fixed point of its beta function.  
 
 \acknowledgments
 The CP${}^3$-Origins centre is partially funded by the Danish National Research Foundation, grant number DNRF90.
KI is supported in part by the DOE grant DE-SC0009919.


\begin{thebibliography}{99}                                                                                               
\bibitem{Litim:2014uca} 
  D.~F.~Litim and F.~Sannino,
  JHEP {\bf 1412}, 178 (2014)
  [arXiv:1406.2337 [hep-th]].



\bibitem{Pica:2010xq} 
  C.~Pica and F.~Sannino,
  Phys.\ Rev.\ D {\bf 83}, 035013 (2011)
  [arXiv:1011.5917 [hep-ph]].



\bibitem{Sannino:2014lxa} 
  F.~Sannino and I.~M.~Shoemaker,
  arXiv:1412.8034 [hep-ph]. To appear in Phys. Rev. D. 


\bibitem{Nielsen:2015una} 
  N.~G.~Nielsen, F.~Sannino and O.~Svendsen,
  Phys.\ Rev.\ D {\bf 91}, 103521 (2015)
  [arXiv:1503.00702 [hep-ph]].


\bibitem{Litim:2015iea} 
  D.~F.~Litim, M.~Mojaza and F.~Sannino,
  arXiv:1501.03061 [hep-th]. 


\bibitem{Rischke:2015mea} 
  D.~H.~Rischke and F.~Sannino,
  arXiv:1505.07828 [hep-th]. To appear in Phys. Rev. D. 


\bibitem{Gross:1973id} 
  D.~J.~Gross and F.~Wilczek,
  Phys.\ Rev.\ Lett.\  {\bf 30}, 1343 (1973).


\bibitem{Politzer:1973fx} 
  H.~D.~Politzer,
  Phys.\ Rev.\ Lett.\  {\bf 30}, 1346 (1973).


\bibitem{Gross:1973ju} 
  D.~J.~Gross and F.~Wilczek,
  Phys.\ Rev.\ D {\bf 8}, 3633 (1973).


\bibitem{Cheng:1973nv} 
  T.~P.~Cheng, E.~Eichten and L.~F.~Li,
  Phys.\ Rev.\ D {\bf 9}, 2259 (1974).


\bibitem{Callaway:1988ya} 
  D.~J.~E.~Callaway,
  Phys.\ Rept.\  {\bf 167}, 241 (1988).


\bibitem{Holdom:2014hla} 
  B.~Holdom, J.~Ren and C.~Zhang,
  JHEP {\bf 1503}, 028 (2015)
  [arXiv:1412.5540 [hep-ph]].


\bibitem{Giudice:2014tma} 
  G.~F.~Giudice, G.~Isidori, A.~Salvio and A.~Strumia,
  JHEP {\bf 1502}, 137 (2015)
  [arXiv:1412.2769 [hep-ph]].
  
  
\bibitem{Seiberg:1994pq} 
  N.~Seiberg,
  Nucl.\ Phys.\ B {\bf 435}, 129 (1995)
  [hep-th/9411149].


\bibitem{Intriligator:1995au} 
  K.~A.~Intriligator and N.~Seiberg,
  Nucl.\ Phys.\ Proc.\ Suppl.\  {\bf 45BC}, 1 (1996)
  [hep-th/9509066].


  

\bibitem{Mack:1975je} 
  G.~Mack,
  Commun.\ Math.\ Phys.\  {\bf 55}, 1 (1977).



\bibitem{Cardy:1988cwa} 
  J.~L.~Cardy,
  Phys.\ Lett.\ B {\bf 215}, 749 (1988).



\bibitem{Osborn:1989td} 
  H.~Osborn,
  Phys.\ Lett.\ B {\bf 222}, 97 (1989).


\bibitem{Jack:1990eb} 
  I.~Jack and H.~Osborn,
  Nucl.\ Phys.\ B {\bf 343}, 647 (1990).



\bibitem{Cappelli:2000dv} 
  A.~Cappelli, G.~D'Appollonio, R.~Guida and N.~Magnoli,
  PoS tmr {\bf 2000}, 005 (2000)
  [hep-th/0009119].


\bibitem{Cappelli:2001pz} 
  A.~Cappelli, R.~Guida and N.~Magnoli,
  Nucl.\ Phys.\ B {\bf 618}, 371 (2001)
  [hep-th/0103237].



\bibitem{Komargodski:2011vj} 
  Z.~Komargodski and A.~Schwimmer,
  JHEP {\bf 1112}, 099 (2011)
  [arXiv:1107.3987 [hep-th]].


\bibitem{Komargodski:2011xv} 
  Z.~Komargodski,
  JHEP {\bf 1207}, 069 (2012)
  [arXiv:1112.4538 [hep-th]].


\bibitem{Anselmi:1997am} 
  D.~Anselmi, D.~Z.~Freedman, M.~T.~Grisaru and A.~A.~Johansen,
  Nucl.\ Phys.\ B {\bf 526}, 543 (1998)
  [hep-th/9708042].


\bibitem{Anselmi:1997ys} 
  D.~Anselmi, J.~Erlich, D.~Z.~Freedman and A.~A.~Johansen,
  Phys.\ Rev.\ D {\bf 57}, 7570 (1998)
  [hep-th/9711035].


\bibitem{Intriligator:2003jj} 
  K.~A.~Intriligator and B.~Wecht,
  Nucl.\ Phys.\ B {\bf 667}, 183 (2003)
  [hep-th/0304128].




\bibitem{Kutasov:2003ux} 
  D.~Kutasov,
  hep-th/0312098.

\bibitem{Kutasov:2003iy} 
  D.~Kutasov, A.~Parnachev and D.~A.~Sahakyan,
  JHEP {\bf 0311}, 013 (2003)
  [hep-th/0308071].

\bibitem{Intriligator:2003mi} 
  K.~A.~Intriligator and B.~Wecht,
  Nucl.\ Phys.\ B {\bf 677}, 223 (2004)
  [hep-th/0309201].

\bibitem{Barnes:2004jj} 
  E.~Barnes, K.~A.~Intriligator, B.~Wecht and J.~Wright,
  Nucl.\ Phys.\ B {\bf 702}, 131 (2004)
  [hep-th/0408156].

\bibitem{Kutasov:2004xu} 
  D.~Kutasov and A.~Schwimmer,
  Nucl.\ Phys.\ B {\bf 702}, 369 (2004)
  [hep-th/0409029].


\bibitem{Banks:1981nn} 
  T.~Banks and A.~Zaks,
  Nucl.\ Phys.\ B {\bf 196}, 189 (1982).


\bibitem{Antipin:2013pya} 
  O.~Antipin, M.~Gillioz, E.~M\o lgaard and F.~Sannino,
  Phys.\ Rev.\ D {\bf 87}, no. 12, 125017 (2013)
  [arXiv:1303.1525 [hep-th]].



\bibitem{Machacek:1983tz} 
  M.~E.~Machacek and M.~T.~Vaughn,
  Nucl.\ Phys.\ B {\bf 222}, 83 (1983).



\bibitem{Machacek:1983fi} 
  M.~E.~Machacek and M.~T.~Vaughn,
  Nucl.\ Phys.\ B {\bf 236}, 221 (1984).


\bibitem{Machacek:1984zw} 
  M.~E.~Machacek and M.~T.~Vaughn,
  Nucl.\ Phys.\ B {\bf 249}, 70 (1985).


\bibitem{Jack:1984vj} 
  I.~Jack and H.~Osborn,
  Nucl.\ Phys.\ B {\bf 249}, 472 (1985).




\bibitem{Hasenfratz:1991it} 
  A.~Hasenfratz, P.~Hasenfratz, K.~Jansen, J.~Kuti and Y.~Shen,
  Nucl.\ Phys.\ B {\bf 365}, 79 (1991).


\bibitem{Gerhold:2007yb} 
  P.~Gerhold and K.~Jansen,
  JHEP {\bf 0709}, 041 (2007)
  [arXiv:0705.2539 [hep-lat]].


\bibitem{Gerhold:2011mx} 
  P.~Gerhold, K.~Jansen and J.~Kallarackal,
  Phys.\ Lett.\ B {\bf 710}, 697 (2012)
  [arXiv:1111.4789 [hep-lat]].


\bibitem{Chu:2015nha} 
  D.~Y.-J.~Chu, K.~Jansen, B.~Knippschild, C.-J.~D.~Lin and A.~Nagy,
  Phys.\ Lett.\ B {\bf 744}, 146 (2015)
  [arXiv:1501.05440 [hep-lat]].

\bibitem{Catterall:2013koa} 
  S.~Catterall and A.~Veernala,
  Phys.\ Rev.\ D {\bf 87}, no. 11, 114507 (2013)
  [arXiv:1303.6187 [hep-lat]].



\bibitem{Ferrara:1974pz} 
  S.~Ferrara and B.~Zumino,
  Nucl.\ Phys.\ B {\bf 87}, 207 (1975).

\bibitem{Grinstein:2008qk} 
  B.~Grinstein, K.~A.~Intriligator and I.~Z.~Rothstein,
  Phys.\ Lett.\ B {\bf 662}, 367 (2008)
  [arXiv:0801.1140 [hep-ph]].


\bibitem{Novikov:1983uc} 
  V.~A.~Novikov, M.~A.~Shifman, A.~I.~Vainshtein and V.~I.~Zakharov,
  Nucl.\ Phys.\ B {\bf 229}, 381 (1983).


\bibitem{Zamolodchikov:1986gt} 
  A.~B.~Zamolodchikov,
  JETP Lett.\  {\bf 43}, 730 (1986)
  [Pisma Zh.\ Eksp.\ Teor.\ Fiz.\  {\bf 43}, 565 (1986)].




\bibitem{Leigh:1995ep} 
  R.~G.~Leigh and M.~J.~Strassler,
  Nucl.\ Phys.\ B {\bf 447}, 95 (1995)
  [hep-th/9503121].

\bibitem{Intriligator:2005if} 
  K.~A.~Intriligator,
  Nucl.\ Phys.\ B {\bf 730}, 239 (2005)
  [hep-th/0509085].

\bibitem{Amariti:2012wc} 
  A.~Amariti and K.~Intriligator,
  JHEP {\bf 1211}, 108 (2012)
  [arXiv:1209.4311 [hep-th]].
  
\bibitem{D'Hoker:2010mk} 
  E.~D'Hoker and Y.~Guo,
  JHEP {\bf 1005}, 088 (2010)
  [arXiv:1001.4808 [hep-th]].
  
\bibitem{Martin:2000cr} 
  S.~P.~Martin and J.~D.~Wells,
  Phys.\ Rev.\ D {\bf 64}, 036010 (2001)
  [hep-ph/0011382].
  
\bibitem{Hofman:2008ar} 
  D.~M.~Hofman and J.~Maldacena,
  JHEP {\bf 0805}, 012 (2008)
  [arXiv:0803.1467 [hep-th]].
  
\end{thebibliography}
\end{document}